# Physiological and Affective Computing through Thermal Imaging: A Survey

YOUNGJUN CHO, Department of Computer Science, University College London (UCL), United Kingdom
NADIA BIANCHI-BERTHOUZE, Division of Psychology and Language Sciences, University College London (UCL), United Kingdom

## ABSTRACT

Thermal imaging-based physiological and affective computing is an emerging research area enabling technologies to monitor our bodily functions and understand psychological and affective needs in a contactless manner. However, up to recently, research has been mainly carried out in very controlled lab settings. As small size and even low-cost versions of thermal video cameras have started to appear on the market, mobile thermal imaging is opening its door to ubiquitous and real-world applications. Here we review the literature on the use of thermal imaging to track changes in physiological cues relevant to affective computing and the technological requirements set so far. In doing so, we aim to establish computational and methodological pipelines from thermal images of the human skin to affective states and outline the research opportunities and challenges to be tackled to make ubiquitous real-life thermal imaging-based affect monitoring a possibility.

## 1 Introduction

Humans are warm-blooded being, self-regulating their own body core and skin temperatures to address environmental changes or internal needs [46]. As this regulation process involves numerous physiological activities, the temperature has acted as a lens to understand the human body and mind. With advances in infrared thermal imaging, a contact-free temperature measurement technique, studies have explored the relation of the skin temperature with other types of physiological activities, psychological and affective states, opening a new pathway to contact-less *physiological and affective computing*. This paper is concerned with this, with the main focus on reviewing and evaluating the rich body of research on thermal imaging-based psychophysiological and affective computing over the past two decades.

In this introduction, we start looking at the past use of earlier thermometry in Section 1.1. Section 1.2 introduces thermal imaging technology and its applications. This is followed by Section 1.3 that introduces mobile thermal imaging that supports real-world physiological and affective computing applications. We also overview specifications of commercially available thermal imaging systems. Finally, we describe the structure of the rest of this paper and our main contributions.

## 1.1 Brief history of human temperature measurement

Back to the eighteenth century when the first mercury thermometer was invented by D.G. Fahrenheit in 1714 [33], physiologists and philosophers had begun to explore bodily heat of a living mammal in relation to medical symptoms [20,92]. Authors observed that certain types of diseases or inflammatory disorders could induce an increase of the temperature of the body. In the nineteenth century, the bodily temperature of patients had been more rigorously investigated; in turn, the heat monitoring had started to be used as a diagnostic and prognostic tool for fatal diseases such as tumors and phthisis [23,61,101]. In the late nineteenth century when physiologists brought up the underexplored topic of the cognitive, sensorial, and emotional functions of the brain to the research community [62], their earlier attentions given to the body temperature had been extended to the heat from the brain. In 1877, to the best of the author's knowledge, Broca applied thermometry to the surface of the head, for the first time, to estimate the mean temperature of the brain [47,123][1]. A few years later, Lombard investigated the head temperature under *cognitive* and *emotional* conditions [79,80].
Explorations on the temperature of the human body had continued throughout the first half of the twentieth century in which thermometry became more accurate and more general in use; in particular, psychiatrists and neurologists had studied the effect of affects and emotions on the bodily temperature of a human being [122] and emotional hypothermia in an animal [42]. These earlier discoveries with the thermometry over the previous centuries paved the way for the understanding of what temperature changes of our body regions would mean; however, it was almost impractical to observe temperatures from the entire skin surficial area which may have different patterns until a new type of thermometry was invented and commercialized around the middle of the twentieth century – that is, *the infrared thermal imaging camera* (see the historical review of the thermal imaging devices; Lloyd's [78]).

## 1.2 Thermal imaging (thermography) technology

Thermal imaging is a non-contact, non-invasive technique to observe heat distributions on the surface of materials and organisms based on the interpretation of naturally emitted electromagnetic radiations into temperatures. Most commercial thermographic cameras sense the electromagnetic radiations (the wavelength range: between approximately 8μm and 14μm - infrared) from surfaces, producing their thermal images, also called thermograms [78]. The ability to read electromagnetic radiations outside the visible spectral range (visible spectrum) differentiates their advantages from the visible spectrum-based imaging devices, such as RGB cameras, which are susceptible to *illumination effects*; for instance, extremes of darkness and brightness incapacitate imaging capabilities due to sensor saturation or sensitivity [3,76].
Thermal images of cutaneous and subcutaneous skin regions (e.g. facial thermogram in Figure 1a) have been used in medical applications to detect pathological symptoms, disorders and diseases of patients. Clinical studies using thermal imaging have demonstrated its high performance in detecting inflammatory arthritis, osteoarthritis, soft tissue rheumatism, and malignant diseases or tumors, and a good performance in clinical monitoring of Complex Regional Pain Syndrome (CRPS) and Raynaud's phenomenon, whose symptoms include abnormal or asymmetric temperature distributions visually inspectable by clinicians [100,110,118].
Thermography has been adopted beyond medical contexts to investigate the possibility of monitoring physiological processes [36,38,75,83,93] and affective states [25,87,90,99,114]. Typically used technology for the monitoring, such as chest belts, probes and electrodes, can be obtrusive, requiring users to wear sensors [5,50,54]. Alternatively, contactless digital image sensor based remote-photoplethysmography (PPG) can be used [37,81,115,120]. However, remote PPG requires an ambient source of light, and does not work properly in dark places or under varying lighting conditions. It also raises privacy concerns (e.g. in hospital). Thermal imaging can be free from those constraints.
The findings of the pioneering research using thermal imaging have drawn upon *high-cost*, heavyweight (several kilograms) thermographic systems as shown in Figure 1b. They require a fixed set up and powerful computing machines for data collection and analysis, limiting their use in highly constrained, systematic laboratory studies. This static thermal imaging setup involves controlled environmental temperature (e.g. stable room temperature) and one's restrained mobility (e.g. a chinrest often used). These requirements had made thermal imaging-based approaches regarded as an infeasible and impractical solution for real world physiological and affective computing applications [37] until the emergence of recently-launched mobile thermal imaging systems.

[1] As Broca's original paper was not available online in 2015-2019 when the survey was written, we referred to two review articles published in 1877 (The Lancet) and in 1985 (The Western Journal of Medicine).

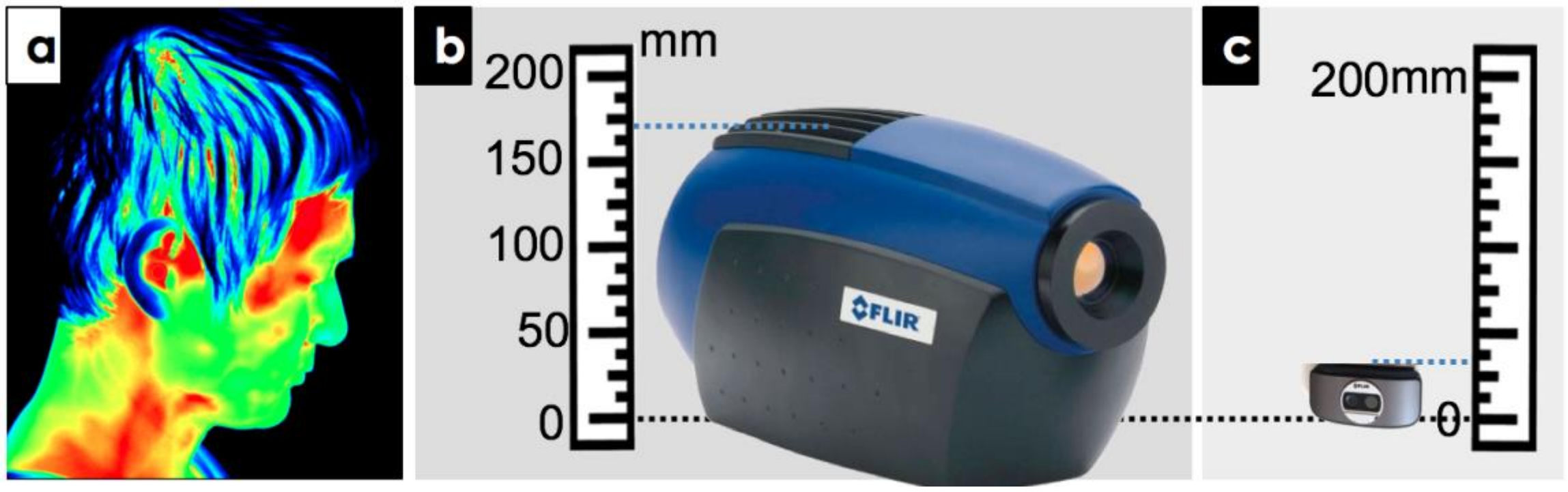

**Figure 1**. (a) an example of a thermal image of the author taken by (b) a high-precision, high-cost, heavyweight thermal camera (FLIR SC5000mb) – dimension: 320x141x159 $mm^3$; weight: 3,800.0g (this camera also needs to be used with a powerful desktop/laptop for data collection). (c) a low-precision, low-cost, lightweight thermal camera (FLIR One 2G) – dimension: 34x68x14 $mm^3$; weight: 36.5g

## 1.3 Mobile thermal imaging

This is an excitement moment for researchers in the field as further advances in thermal technology have being made producing a new category of portable thermographic systems: mobile, low-cost thermal imaging devices (e.g. Figure 1c – price: lower than £200, weight: 36.5g, dimension: 34x68x14 $mm^3$). This mobile technology could help bridge the gap between the findings from highly constrained laboratory environments and in the wild real-world applications. Indeed, recent studies have started to explore *mobile thermal imaging* to support unconstrained, real world use cases of thermal imaging-based physiological monitoring [14,16], affect recognition [12,15] and context monitoring [13].

Mobile thermal imaging can provide high levels of flexibility and scalability for recovering physiological signatures and recognizing a person's affective states. Users can then use the system ad-hoc in situations when they need to monitor their condition as shown in Figure 2. A person can just hold a smartphone facing his/her face whilst walking to monitor breathing patterns (Figure 2a). This flexible setup enables many other uses, such as attaching the system on a desktop, desk, handle of the bicycle, at a dashboard or even on a window or mirror in the car, similar to set-ups used in other studies [17,49] – for example, it can be mounted on a chair in Figure 2b and installed on an eyeglass frame [40]. This flexible setup can help thermographic cameras to capture physiological thermal signatures of interest effectively, whilst at the same time they do not interfere too much with a person's view. In a recent study on mental stress monitoring of a worker [14], a low-cost thermal camera was mounted on a mixed reality head-mounted display (Hololens) as shown in Figure 2c. With this setup, the worker could be informed of detected stress levels to help tailor work schedules to his/her psychological needs.

Also, mobile thermal imaging can remove the need to wear sensors (e.g. breathing belt) that may not be suitable or unpleasant. For example, in the gym people may find sensors unpleasant due to sweat or constrained movements. Another case would be people with a certain type of pain such as CRPS [7] who tend to reduce the number of clothes and objects that touch their skin. Given this, mobile thermal imaging could drive its scalability in real world settings.

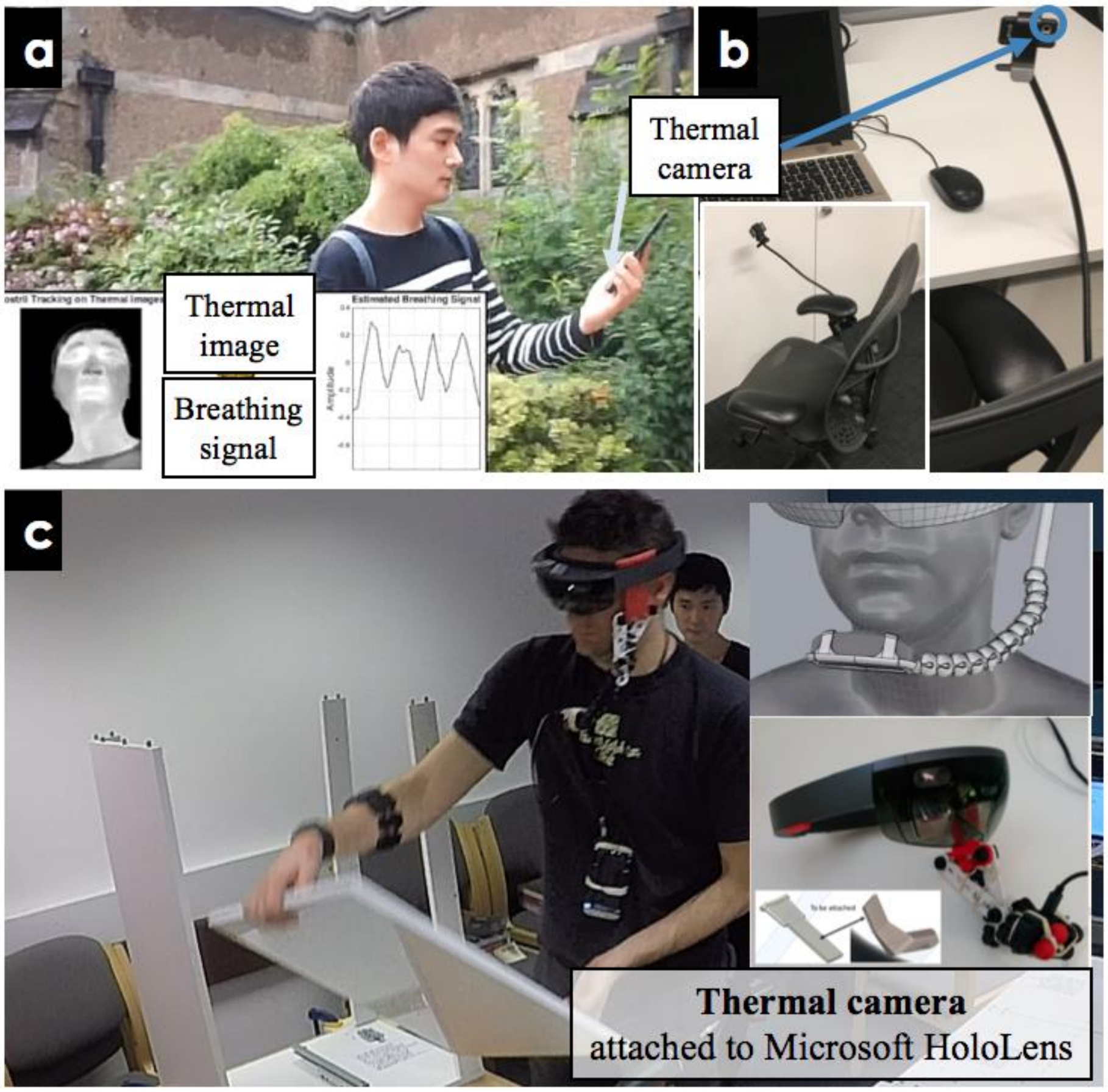

**Figure 2.** Use cases of mobile thermal imaging-based physiological and affective computing: a) breathing monitoring with a low-cost thermal camera attached on a smartphone [16], b) a hardware frame attached to a chair for monitoring a person's upper body and face, c) stress monitoring of a worker in a manufacturing shop floor. A low-cost thermal camera was attached to an MR headset (Microsoft HoloLens) [14].

### Information of commercially available thermal cameras (in 2019)

Table 1 summarizes specifications and prices of both low-cost and high-cost thermal imaging devices available between 2015 and 2019. There are three key parameters in determining thermal imaging quality: thermal sensitivity, spatial resolution and temporal resolution (sampling rate). The thermal sensitivity is measured by Noise Equivalent Temperature Difference (NETD). NETD describes the minimal temperature difference that is recognizable by a thermal imaging device. Hence, the smaller, the better. The NETD of thermal cameras varies from 0.017°C to 0.5°C. The cost of the currently available sensors seems to primarily depend on their temporal resolution (here called sampling rate). For example, while both FLIR One 2G (at $239, in May 2016) and Optris PI 200 (at $4,250, in August 2019) meets the 160x120 spatial resolution and under 0.1°C sensitivity, the former has unsteady sampling rate lower than 8.7Hz and the latter supports 128Hz. Nonetheless, such low-cost cameras (+ in Table 1) provide valuable benefits, such as small form factors, affordability and portability (e.g. lower than 34x14x68mm$^3$ dimension), making them more feasible to be deployed in HCI systems than heavyweight, immobile, and expensive high-end thermographic systems (* in Table 1).

**Table 1.** Specification and price of commercially available thermal cameras

| Product | Spatial Resolution | Sampling Rate | Thermal Sensitivity (NETD) | Device Dimension | Weight | Price± (in August 2019) |
|---|---|---|---|---|---|---|
| +FLIR One 3G | 80x60 | <8.7Hz (unsteady) | 0.10℃** | 68 x 34 x 14 mm$^3$ | 34.5g | $199.99 |
| +FLIR One 2G | 160x120 | <8.7Hz (unsteady) | <0.10℃ | 68 x 34 x 14 mm$^3$ | 36.5g | Not available (in Aug 2019)<br><br>**£166.00** (about $239 in May 2016) |
| +Seek Thermal Compact | 206x156 | <9Hz (unsteady) | 0.50℃ | 25.4 x 45 x 20 mm$^3$ | 14g | $249.00 |
| +Seek Thermal Compact XR | 206x156 | <9Hz (unsteady) | Unknown | 25.4 x 45 x 25.4 mm$^3$ | 14g | $299.00 |
| +FLIR One Pro | 160x120 | <8.7Hz (unsteady) | 0.07℃** | 68 x 34 x 14 mm$^3$ | 36.5g | $399.99 |
| +Seek Thermal Compact PRO | 320x240 | <9Hz (unsteady) | <0.07℃ | 25.4 x 45 x 25.4 mm$^3$ | 14g | $499.00 |
| +Therm-App | 384x288 | 8.7Hz | <0.07℃ | 55 x 65 x 40 mm$^3$ | 138g | $999.00 |
| +Therm-App Hz | 384x288 | 25Hz | <0.07℃ | 55 x 65 x 40 mm$^3$ | 138g | $1,700.00 |
| +Therm-App PRO 640 | 640x480 | 25Hz | <0.03℃ | 55 x 65 x 40 mm$^3$ | 138g | $3,999.00 |
| *Optris PI 200 | 160x120 | 128Hz | <0.10℃ | 90 x 45 x 45 mm$^3$ | 215g | $4,250.00 |
| *FLIR A35 | 320x256 | 60Hz | <0.05℃ | 106 x 40 x 43 mm$^3$ | 200g | $5,900.00 |
| *OPTRIS PI-640 G7 | 640x480 | 125Hz | 0.130℃ | 100 x 56 x 46 mm$^3$ (Without lens) | 320g | $9,995.00 |
| *FLIR A325sc | 320x240 | 60Hz | <0.05℃ | 170 x 70 x 70 mm$^3$ (Without lens) | 700g | $13,216.33 |
| *FLIR A655sc | 640x480 | 50Hz | <0.03℃ | 216 x 73 x 75 mm$^3$ (Without lens) | 900g | $24,662.43 |
| *FLIR SC5000mb | 640x512 | 100Hz | 0.017℃ | 320x141x159 mm$^3$ | 3800g | Unknown |
| *FLIR SC7650 | 640x512 | 100Hz | 0.020℃ | 253x130x168 mm$^3$ (Without lens) | 4950g (Without lens) | Unknown |

+ A mobile phone is needed for data collection
* A powerful desktop/laptop is needed
** The manufacturer corrected the thermal sensitivity values on datasheets in 2018 (from 0.15℃).
± Price for software is not included

## 1.4 Article Organization and Contributions

In this survey, we dive into existing methods, paradigms and physiological evidence around the thermal responses in association with physiological and affective states. Firstly, we discuss how thermal imaging can be used to measure physiological signatures and identify earlier challenges. Secondly, we review studies exploring the use of thermal imaging in monitoring a person’s affective states and discuss their findings. Finally, we state the challenges and limitations which have emerged from the literature.

The main contributions of this article are three-fold. First, we establish computational and methodological pipelines from thermal images of the human skin to physiological signatures and affective states. Second, we provide a guideline of contactless physiological and affective computing through thermal imaging, including how to measure, system requirements and how to evaluate methods. Finally, we highlight future research opportunities and directions given challenges we have identified in terms of where we are and how to move ourselves forward.

# 2 Physiological thermal signatures

In this section, we review existing works about the use of thermal imaging in measuring a person's physiological cues that are considered related to affective states. We are particularly interested in i) what types of physiological thermal signatures have been explored, ii) which computational methods and metrics have been developed, and iii) the robustness of related findings.

## 2.1 Computational pipeline: an overview

Before diving into each physiological thermal signature, we establish a computational pipeline consisting of common technical components emerging from the literature on thermal imaging-based physiological computing. The pipeline starts with the selection of a region-of-interest (ROI) on thermal images of the human skin as shown in Figure 3. Different types of physiological thermal signatures can be extracted by monitoring temperatures on different ROIs. For example, nostrils can be chosen as a ROI for the monitoring of the respiratory activity (e.g. breathing rate) [16]. The ROI can be automatically tracked on sequential thermal images (also called a thermal video) using an extra physical marker (e.g. dot stick) [21,86] or an advanced computer vision algorithm [16,90]. At the final stage of the pipeline, temperature patterns on the tracked ROIs have to be interpreted for producing a physiological timeseries signal (e.g. respiratory signal in Figure 3). For the interpretation, the averaging has been dominantly used [1,2,35,75,83] given its simple computation and minimal cost [104].

For the automatic ROI tracking, thermal images must be quantized or converted into a digital signal with a fixed number of bits because a thermal camera captures a matrix of temperature, not an image directly. This step is called *quantization* (Figure 3a). The most common approach is to use linear quantization with a selected temperature range of interest, which is traditionally fixed from the first thermogram frame (e.g. 28°C to 38°C in [34,84]). However, this is unable to adapt to dynamic situations where environmental temperature varies [16]. These factors directly influence automatic ROI tracking performance. To tackle this, a recent study has proposed the optimal quantization (OQ) method that adaptively maps temperatures to thermal images against environmental temperature effects [16] (for details, see Section 2.3).

Automatic ROI tracking on thermal images, however, is complex to implement and there is no standard computational method supporting this. Also, the focus of a variety of studies in this field has been on capturing binary directions in temperature changes on a ROI occurring in association with a person's affective state (see next sections for the details). In this case, it does not necessarily require a support from complex computational methods for the automatic ROI tracking as shown in Figure 3b. For example, Engert *et al.* [30] visually inspected and manually rejected thermal images where there is a movement of a participant on the images. To reduce this effort, other authors have requested their participants to stay still during the task [44,103]. A head fixation mount, such as a chinrest (Figure 3b), has been often used to make this process easier [63,114].

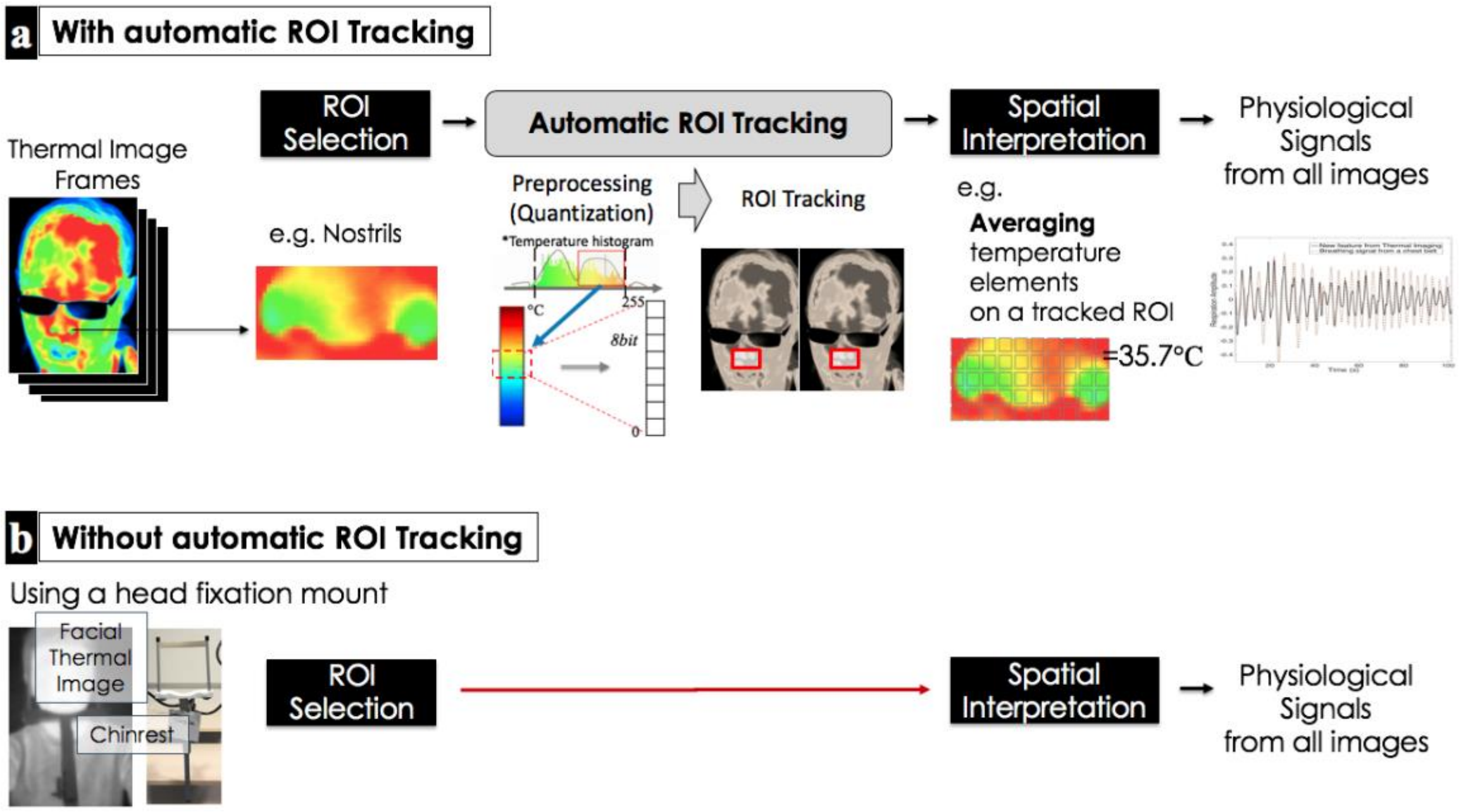

**Figure 3.** The established computational pipeline that has been commonly applied to studies on thermal imaging-based physiological computing: this consists of three main steps, the ROI selection, automatic ROI tracking and spatial interpretation. a) with automatic ROI tracking, b) without automatic ROI tracking (in this case a head fixation mount is used).

## 2.2 Cardiovascular thermal signature

The most widely explored physiological thermal signature is thermal directional change on the skin [21,30,39,87,103]. Blood flow makes heat transferred from the body core to the skin. The heat conduction is regulated by vasoconstriction which is controlled mainly by the sympathetic nervous system [46,116]. Vasoconstriction induces a local decrease in temperature. This is the narrowing of blood vessels, causing blood flow (to the skin) to decrease and in turn reducing loss of body core heat. Hence, the surface area becomes colder. In an healthy body, this can occur under cold ambient temperature conditions [94] as well as a specific affective state, such as mental distress [28]. Vasodilation has the opposite effect. Given this regulatory phenomenon, most researchers in this domain have focused on studying the relationship between thermal directional changes (i.e. temperature drop, rise) of specific skin areas in relation to psychophysiological states (e.g. stress) as discussed in Section 3.1. In particular, the nose tip has been the main facial area investigated [2,14,21,30,39,87,103,114] in relation to vasoconstriction and vasodilation phenomena.

As mentioned in the previous section, the averaging of temperatures on the skin ROI has been used to convert spatial temperature patterns into the cardiovascular signals (e.g. [30,87,114]). With the average of temperature, the thermal directional change as a physiological metric can be calculated in three ways: i) comparison of the average temperature computed at the beginning of a task with the one computed at the end of the task [14,87], ii) difference between the average temperatures between tasks computed either at the end of the task or average across a period of it [2,21,87,103,114], and iii) computation of the linear slope characterizing the temperature measurement sequence during the duration of the task [14,30]. Given that the main focus of the body of work has been to observe the thermal directional changes due to specific tasks considered inducing psychological states (see Section 3.1), there has not been a direct evaluation method to test the quality of the physiological signatures and metrics.

As these metrics are of relative ease to compute, it has been unnecessary to use complicated automatic ROI tracking methods to monitor the physiological signals continuously and automatically [21,30,87,103,114]. Although a few studies used automated motion tracking software, the purpose was to correct minor motion artefacts to the difficulty of participants to remain fully still [2]. A critical issue when using a thermal camera is their sensitivity to thermal environmental changes. Hence, to minimize or remove environmental temperature effects during experiments, room temperatures have been generally controlled [2,39,87]. Although efforts have been made recently to address challenges of both motion artefact and environmental temperature changes with mobile thermal imaging [14,16], such constraints were prerequisite for most of studies where heavy thermographic systems (i.e. static thermal imaging) were used in the literature reviewed in the rest of this paper.

Another type of cardiovascular thermal signatures investigated in the literature is the cardiac pulse. Given that the temperature of the skin above a superficial blood vessel is influenced by the blood flowing through it, a few of studies have focused on extracting the pulse rate from thermal images. For example, Garbey et al. [36] analyzed temperatures over a blood vessel area of the neck in the frequency domain for estimating the pulse rate. This work was followed by two studies focusing on signal optimization [8] and filtering [48] with the same underlying principle with [36]. However, all these studies have a fundamental problem in their performance. Despite these authors claiming the high accuracy in measuring an average purse rate (PR in bpm), they used a nonstandard evaluation metric called *Complement of the Absolute Normalized Difference* (CAND) which always produces an extremely high value. For example, the latest work [48] (in 2016) reported the accuracy of 92.46% from the CAND. We evaluated their results (i.e. measured HR data with reference PPG signals in Table 1 in [48]) with the Pearson correlation coefficient, which is one of the standard evaluation metrics in the heartrate monitoring. The result was r=0.58 which can be considered extremely low given the performances reported in other PPG studies (e.g. a remote PPG study reported r=1.00 [97]). Furthermore, by contrast with other types of cardiac pulse measurements (e.g. ECG or PPG), this type of method is unable to provide a rich set of physiological metrics including heart rate variability (HRV) related ones [105] given the low signal quality.

## 2.3 Perspiratory thermal signature

Thermal imaging of a skin area with a high density of sweat glands, such as the palm and perinasal regions, have been used to monitor perspiratory responses (i.e. sweat gland activations). The sweat gland activation and deactivation result in a change in temperature [69,90,108]. Pavlidis *et al*. [90] proposed to compute the instantaneous energy of a spatial temperature pattern on paranasal and finger areas to monitor perspiratory thermal signatures. The authors applied a low-pass filter to the energy signal from the perinasal area to minimize breathing effects on the signals. It is noteworthy that authors found strong correlations between galvanic skin response (GSR) signals from standard electrodermal activity (EDA) sensors and the extracted signals from thermal images of the perinasal area and of a finger region (r=0.943, r=0.968, respectively). As shown in Figure 4, high-resolution thermal cameras (e.g. 1280x1024) with specialized focal lens were required for the monitoring.

More recently, Krzywicki *et al*. [69] proposed a computational method to observe active pores on the cutaneous skin directly using high-resolution thermal imaging. Authors proposed a new physiological metric: pore activation index (PAI) which can be calculated based on the number of counted activated pores. Despite their reported lower correlation coefficient (r=0.71) between PAI transient response and skin conductance response from a finger EDA sensor, the method provides a more direct lens to a person's sweat glands. Also, authors confirmed that fingers are more responsive than facial areas in perspiration. Recently, the monitoring of active pores has been adopted in a human computer interaction application [4].

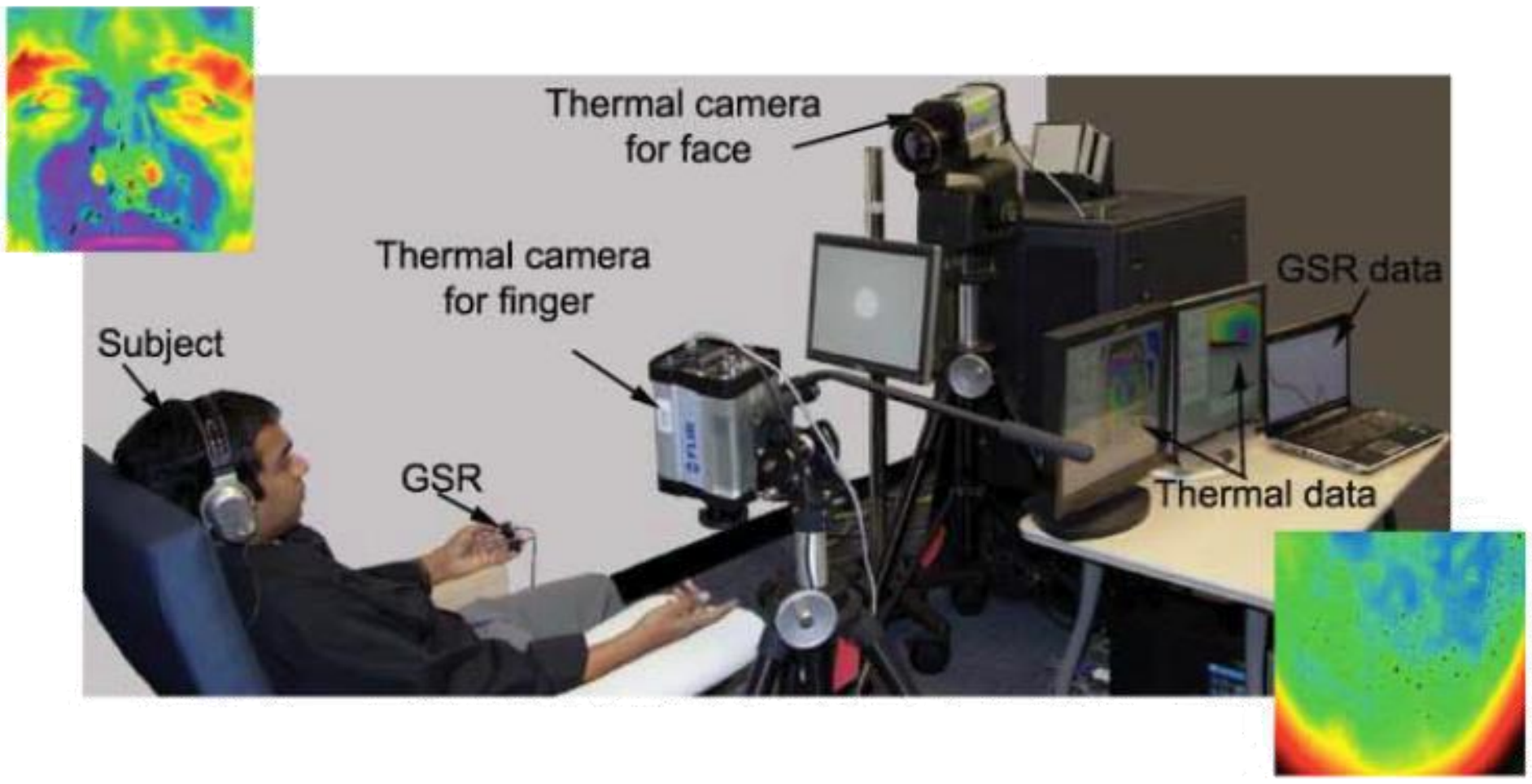

**Figure 4.** A typical static thermal imaging setup with high-resolution thermal cameras for the monitoring of perspiratory thermal signatures (adapted from [90]).

## 2.4 Respiratory thermal signature

Likewise, it has been shown that respiratory responses can be monitored through thermal imaging [1,34,75,83,84,93]. This can be done by monitoring the thermal changes around the nostrils or mouth areas caused by the inhalation and exhalation breathing cycle. Although both the mouth and the nostrils area can be used for this purpose, most of the existing works have chosen the nostril(s) as their key ROI. Like other studies, the simple averaging of temperatures on the ROI has been used to represent the breathing signal. This approach has been tested in several indoor stationary contexts, including neonatal care [1], sleep monitoring [34] and driver's drowsiness monitoring [66], with the thermal camera set in a fixed position.

The quality of the measurement has been mainly evaluated with a belt-type breathing sensor [34,83,84,93]. Earlier works focused on measuring an average breathing rate from a person over the course of time (e.g. 5 mins) [1,34,83,84]. For the evaluation, nonstandard evaluation metrics (e.g. CAND as discussed in Section 2.2) were used. The reported performances in measuring average (or dominant) breathing rates have shown to be much higher than the ones reported for cardiac pulse rate. Later, computational physiologists have focused on more sophisticated respiratory metrics: relative tidal volume, inter-breath interval (IBI) [75] and sequential respiratory rates (RR) using a short time window (e.g. 20s) [16,93]. These require the high quality of breathing timeseries data.

The challenge in tracking high quality breathing signals lies in the difficulty in tracking the nostril area (in comparison to others) even in the case of small movements of the person. Respiration monitoring accuracy has been gradually improving due to the adoption of advanced ROI-tracking methods, for example, spatial representation [82] in [93]. Within indoor controlled experiments, the work reported in [74] achieved strong correlations of sequential respiratory rates (not average rate) with the ground truth (r=0.974) obtained from piezo plethysmography instrument.

Beyond controlled laboratory settings, however, the only use of advanced ROI-tracking methods does not lead to robustness of this measurement. One main challenge has been how to tackle environmental thermal dynamics [16]. For example, Figure 5a shows that variations in ambient temperature can significantly alter facial thermal images [16], decreasing the ROI tracking performances and consequently the breathing monitoring accuracy. Such challenges are similar to tone-mapping-related quantization issues in computer vision when converting real-world luminance to digitally visualized color value [74]. As mentioned earlier, earlier works on thermal imaging have completely ignored the issue by adopting a fixed range of temperature for the quantization (e.g. the range between 28°C and 38°C in [34]).

More recently, Cho et al. [16] have shown that the adverse effects of environmental temperature variations on thermal imaging can be handled by the optimal quantization. Furthermore, the authors have highlighted a fundamental problem of the averaging method for the spatial interpretation (Figure 3a). The averaging method tends to ignore a fairly weak respiratory signature on a thermal image, such as the one during shallow breathing (Figure 5b, 5c). Whilst inhalation and exhalation patterns are distinctive during deep breathing, these patterns are very similar during shallow breathing. The authors have also proposed a new spatial interpretation method that helps integrate thermal voxels. With advanced tracking methods proposed by the same authors, the work has shown extremely strong robustness in monitoring sequential respiratory rates (analogous to IBI) reaching high correlation (r = 0.9987) with a standard respiration-belt. This was also the first work to test a thermal camera based breathing monitoring system in highly dynamic thermal situations during ubiquitous tasks and using a low-cost thermal camera (e.g. walking in the park, climbing stairs) [11,16].

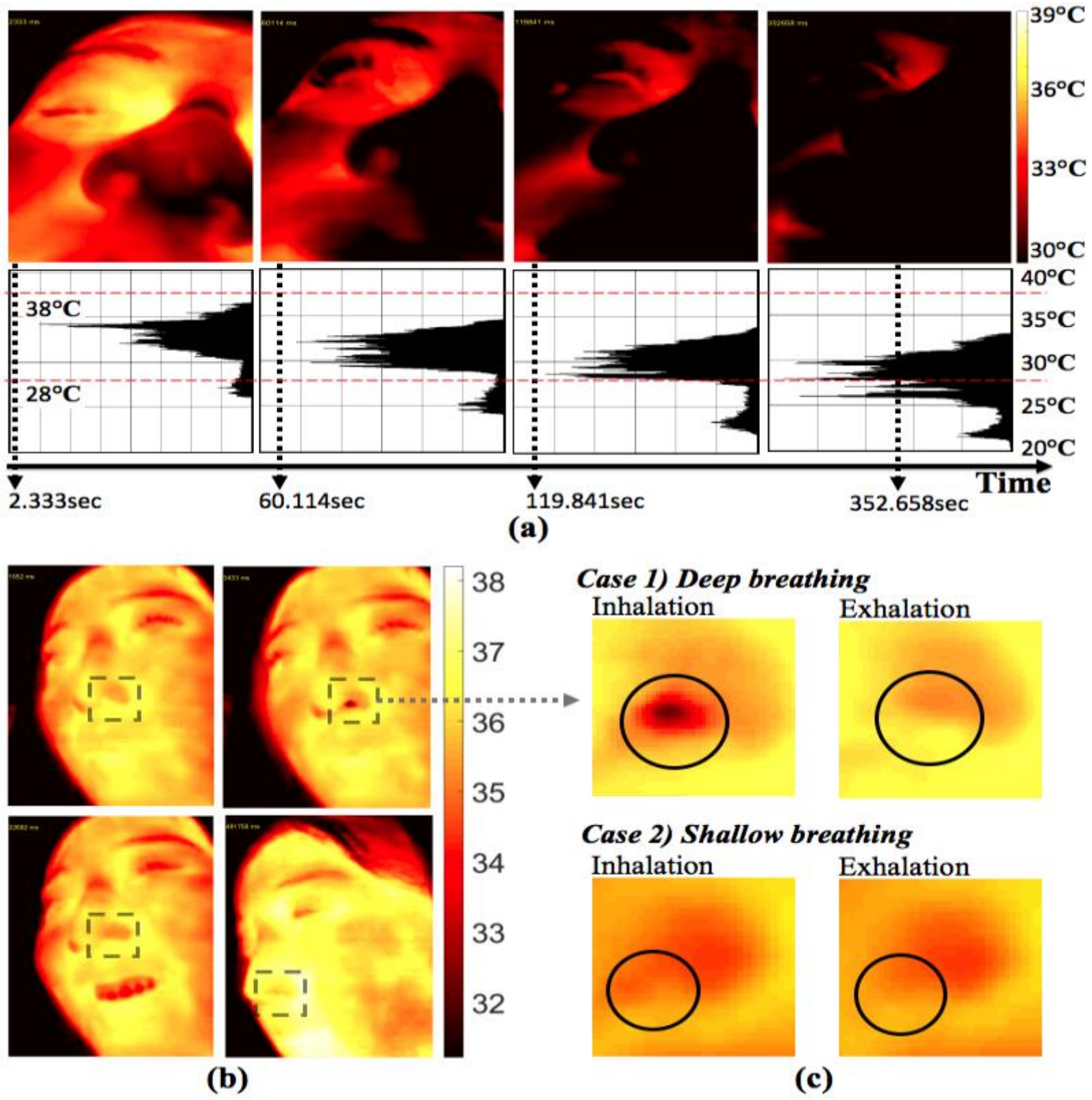

**Figure 5.** Key challenges in thermal imaging-based physiological computing in real world settings (adapted from [16]): (a) examples of thermograms collected from a person walking outdoor for 6 minutes and their histograms, (b) thermal images of a person moving and (c) the quality of respiratory thermal signatures being affected by a breathing type.

## 2.5 Muscular thermal signature

Cues of muscular activations can also be observed through thermal imaging. Facial expressions, for example, are the results of the activation of a number of micro-muscle units [27], resulting in changes in their temperature [63] and possibly also in the alteration of blood flow in certain areas of the face.

To observe facial micro-muscle activations more directly, the facial action coding system (FACS) and facial electromyography (EMG) can be used. In particular, the FACS is based on the manual analysis of anatomical composition of human facial muscle action units (AUs) [26]. While this is very accurate, this requires a trained human coder. Hence, researchers have tried to automatize FACS coders' time-consuming decoding works [63,119].
To the best of the author's knowledge, Jarlier *et al.* [63] first explored temperature patterns on facial AUs while trained FACS coders activated specific AUs to make emotional expressions intentionally. They observed not only the facial temperature distribution, but also the speed and intensity of corresponding muscle contractions. While the participants' heads were immobilized to minimize motion artefacts and the accuracy was still limited, the authors demonstrated potentials of thermography in the FACS decoding process. Wesley *et al.* [119] followed this work by comparing thermal imaging-based FACS decoding with a RGB-vision based one. The authors reported that thermography could be more effective given that facial thermal images are much less affected by ambient lighting conditions, the main limitation in the RGB-vision.
In order to improve the automatic recognition performances, researchers have explored the possibility of combining both the thermal and RGB cameras [5, 40] given the already outstanding advances in the use of the RGB camera for this task [18,106,107]. Later, researchers have also explored the use of thermal imaging alone to this purpose [52,77,117].
Similar to the RGB-vision based approaches, the body of works have focused on a few facial ROIs (i.e. mostly three to five): the eyes, nose, mouth, cheeks and forehead [52,77,117]. Differently from the research carried out on the other physiological signatures discussed in the previous section, the work in this area has gone further than investigating the phenomenon from a thermal imaging perspective. Possibly building on the work on RGB-based facial expression recognition, such studies have adopted a variety of machine learning techniques, such as hidden Markov models in [77] and principal component analysis in [117]so as to improve the recognition performances.

Finally, Table 2 and Figure 6 summarize the literature reviewed in this section. In Figure 6, thickness of a line is to show what a majority of existing works have focused on. As discussed above, the selection of ROIs is linked to the physiological signatures of interest. Amongst explored physiological thermal signatures, vasoconstriction driven thermal directional changes have been mainly explored in relation to affective states which are reviewed in the following section. The most of works have been conducted in systematic constrained settings in terms of environmental temperature changes and movements.

**Table 2.** A summary table of the literature on thermal imaging-based physiological computing.

| **Authors** | **Physiological Signature** | **Participants / activities** | **Automatic ROI Tracking** | **Physiological Metrics** | **Evaluation & Results** | |
|---|---|---|---|---|---|---|
| | | | | | Reference signals | Evaluation metrics |
| **Veltman and Vos (2005)** | Cardiovascular signature (vasoconstriction driven nose-tip temperature change) | 8 healthy participants / memory task | Not used | Temperature difference | Not used* | |
| **Or and Duffy (2007)** | | 33 healthy participants / car-driving simulation | Not used | Temperature difference | | |
| **Di Giacinto et al. (2014)** | | 10 PTSD patients, 10 healthy participants / fear task | Not used (a physical marker used) | Temperature difference | | |
| **Engert et al. (2014)** | | 15 male participants / stress induction tasks | Not used | Temperature slope | | |
| **Salazar-Lopez et al. (2015)** | | 120 healthy participants / IAPS task | Not used | Temperature difference | | |
| **Abdelrahman et al. (2017)** | | 24 healthy participants / Stroop test | Bespoke tracker using OpenCV library | Temperature difference | | |
| **Cho et al. (2019)** | | 25 participants / sitting (10), stress induction task (12), physical assembly task (3) | Thermal gradient flow (TGF) [16] | Temperature difference, Slope, SDSTV, SDTV | | |
| **Garbey et al. (2007)** | Cardiovascular signature (cardiac pulse) | 34 healthy participants / resting in an armchair | Conditional density propagation tracker [60] | Average pulse rate (bpm) per participant | Piezo electric pulse transducer | Non-standard CAND =88.52% (not reliable)** |
| **Hamedani et al. (2016)** | | 22 healthy participants / resting | TLD Tracker [64] | Average pulse rate (bpm) per participant | PPG sensor | Non-standard CAND =92.47% (not reliable)** |
| **Pavlidis et al. (2012)** | Perspiration | 17 surgeons / laparoscopic drill test | Tissue tracker [121] | GSR timeseries signal | Finger EDA sensor | Correlation r=0.968 (finger) r=0.943 (perinasal) |
| **Krzywicki et al. (2014)** | | 20 healthy participants / respiration exercise | Not reported | Pore activation index (PAI) (equivalent to skin conductance response) | Finger EDA sensor | Correlation r=0.71 |
| **Murthy et al. (2004)** | Respiration | 10 healthy participants / resting | Not reported | Average breathing rate per participant | Respiration-belt sensor | Accuracy ($\left\|\frac{estimated}{reference}\right\|$) = 92% |
| **Murthy et al. (2006)** | | 3 healthy participants / resting | Bespoke tracker | Average breathing rate per participant | Respiration-belt sensor | Accuracy ($\left\|\frac{estimated}{reference}\right\|$) = 96.43% |
| **Fei and Pavlidis (2010)** | | 20 healthy participants / resting | Coalitional tracking algorithm [24] | A breathing rate (over 3 mins) per participant | Respiration-belt sensor | CAND* =98.27% |
| **Abbas et al. (2011)** | | 7 infants / no task | Not reported | Average breathing rate (over 5 | ECG belt | Not properly evaluated |
| | | | | mins) per participant | | |
| **Lewis et al. (2011)** | | 12 participants / sedentary breathing exercise | PBVD tracker [45] | Dominant breathing rate (over 2 mins), Inter-breath interval (IBI), relative tidal volume (rTV) | Inductance Plethysmo-graphy | Correlation r=0.941 (dominant breathing rate for normal condition) r=0.98 (IBI) r=0.90 (rTV) |
| **Pereira et al. (2015)** | | 11 healthy participants / sedentary breathing exercise | Mei's tracker [82] | **Sequential** respiratory rates (analogous to IBI) | Piezo plethysmo-graphy | Correlation r=0.974, Mean absolute BR error = 0.96 |
| **Cho et al. (2017)** | | 23 healthy participants / indoor and outdoor breathing exercise, physical activities | Thermal gradient flow (TGF) [16] | **Sequential** respiratory rates (analogous to IBI) | Respiration-belt sensor | Correlation r=0.9987, RMSE=0.46 bpm, Respiratory signal quality index rSQI=0.98<br><br>(N=886 for indoor-outdoor breathing exercise) |
| **Jarlier et al. (2011)** | Muscular thermal signature (facial muscle action unit) | 4 FACS coders / facial expression elicitation tasks | Not used | Speed and intensity of AUs | Not used | Not evaluated |
| **Wesley et al. (2012)** | | 8 healthy participants / facial expression elicitation task (a FACS encoder trained them) | Not used | Mean temperature on an AU. | Not applicable (the focus was on the facial expression classification task) | |

* Given that the main focus of the body of work on vasoconstriction-related cardiovascular signature has been to observe the thermal directional changes due to specific tasks considered inducing psychological states (see Section 3.1), there has not been a direct evaluation method to test the quality of the physiological signatures and metrics.

** As reported in Section 2.2, this nonstandard evaluation metric (CAND) has a fundamental issue (this produces extremely a higher value than standard metrics).

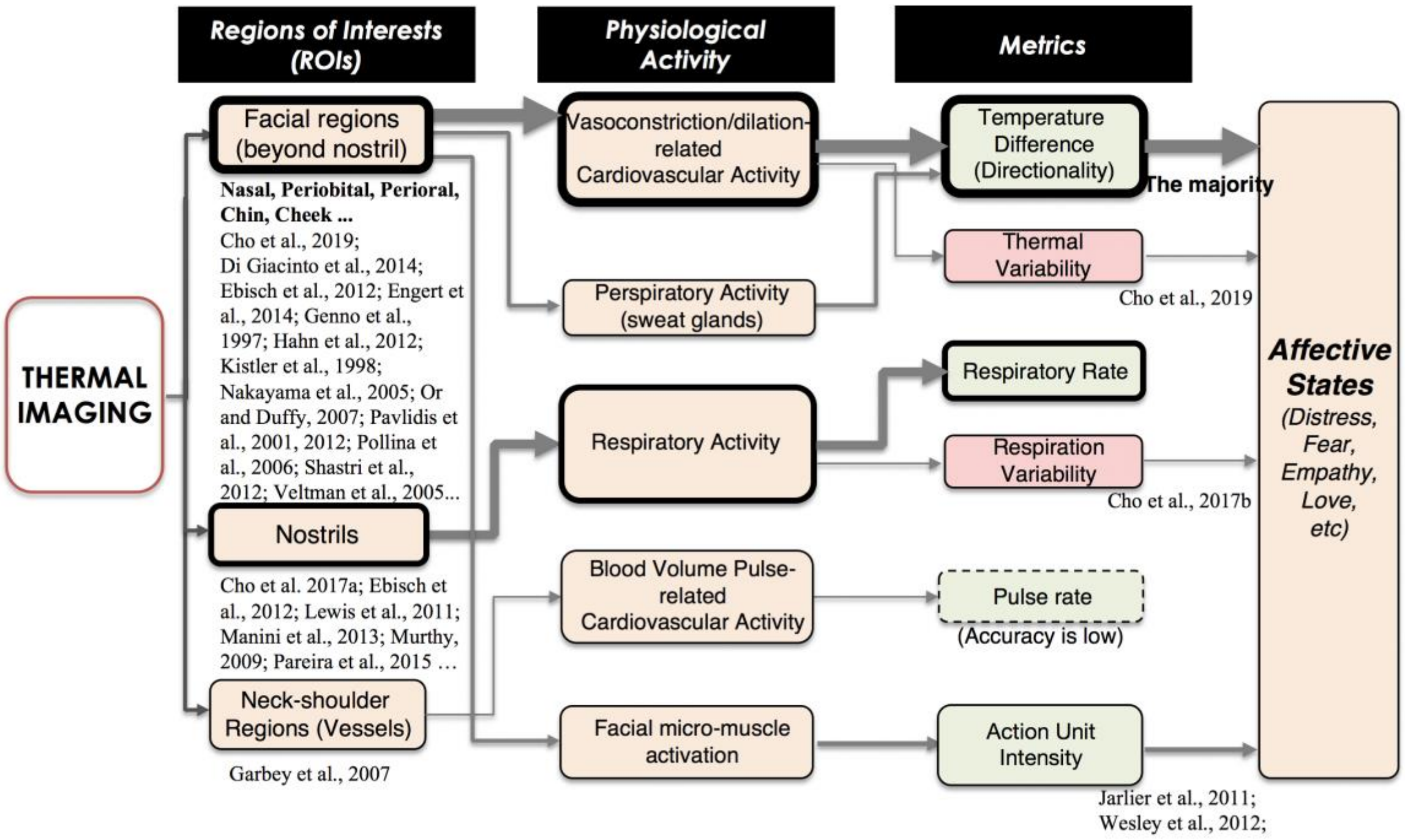

**Figure 6.** A summary diagram of the literature on thermal imaging-based physiological and affective computing: choosing a ROI corresponding to each physiological signature and a metric are key in assessing a person's affective states (the thickness of the lines relates to the amount of literature in that area).

# 3 Thermal imaging as a measure of affective states

With the literature reviewed in Section 2, this section aims to bring together the relevant literature that explores the ability of thermal imaging in assessing a person's affective states and experimental protocols. Before starting, it is noteworthy that while the thermal directional changes of facial regions have been dominantly investigated so far in association with a person's affective states, the roles of other types of physiological thermal signatures in affective computing have been much less explored in this body of work.

## 3.1 Thermal directional change and affective states

Amongst physiological thermal signatures, the vasoconstriction/dilation-related cardiovascular and sweat gland activation-related perspiratory responses induce increases or decreases in temperatures of ROIs, which could be quantified with a simple metric such as the temperature difference between data at two temporal points. Hence, the majority of studies have focused on thermal directional changes on ROIs in association with a person's affective states. The body of work have mostly chosen local areas on facial thermal images - an example of facial thermal image is shown in Figure 7 Left. Studies have investigated the relationship between thermal directional changes and affective states ranging from mental stress and fear to sexual arousal and maternal empathy in contexts of social interaction. Before diving into existing works, Figure 7 shows examples of thermal directional responses to affective states. Below are the detailed reviews of the literature.

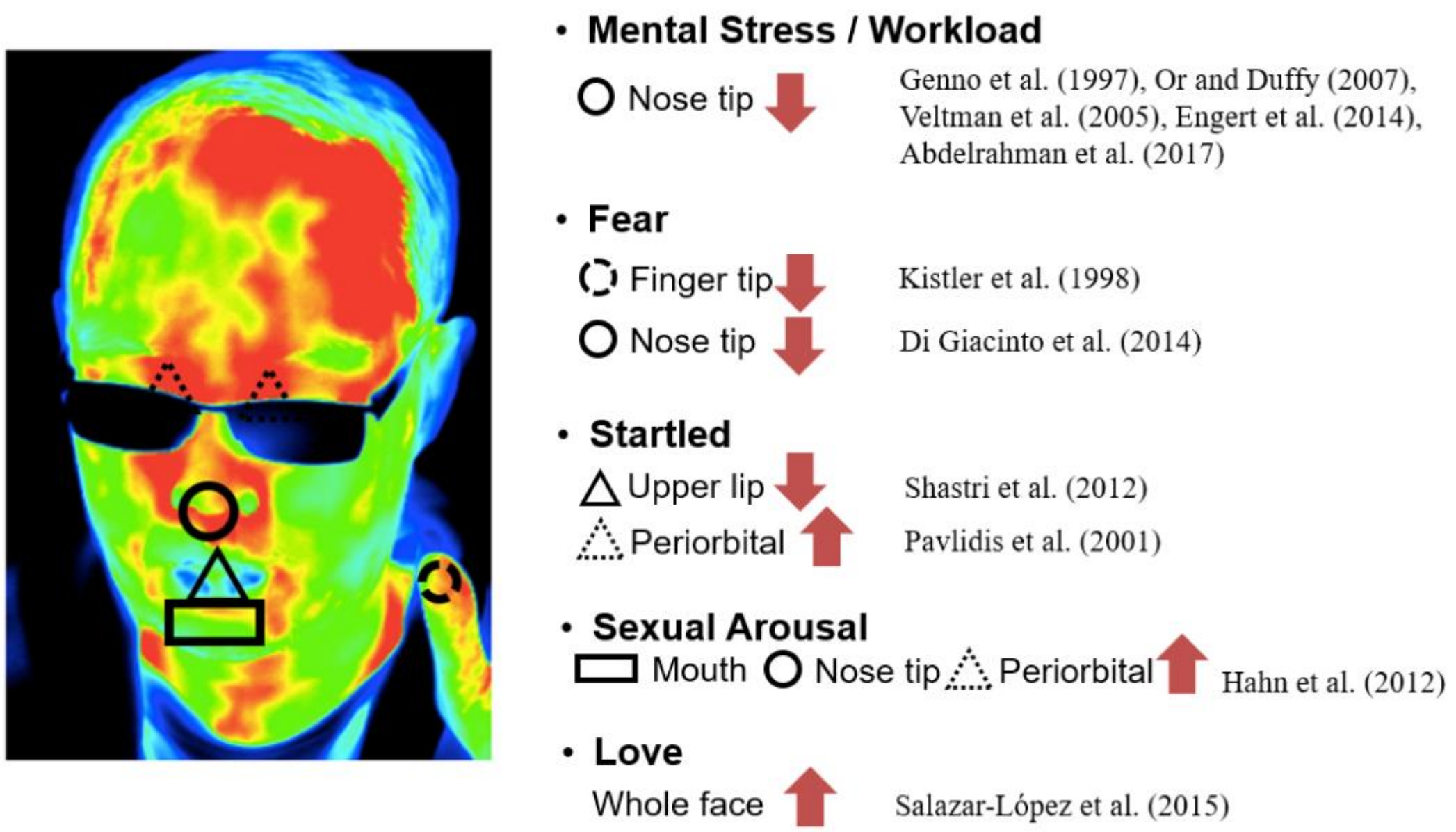

**Figure 7.** Left) an example of thermal image of his face and thumb finger, and widely used ROIs, Right) examples of thermal directional responses of selected ROIs to a person's affective states.

#### Mental stress and workload

Genno *et al.* [39] analyzed nose and forehead temperatures to observe their directional responses to mental stressors. Using a contact-based multi-channel thermistor, they measured temperatures during both a resting period and a self-designed stress induction task. During the tasks, participants were required to keep searching for a moving virtual object on a computer screen which suddenly pops up with a flash lightening effect and a loud siren sound feedback as the onset of stressors. To minimize effects of ambient temperature the authors controlled the room temperature. In the analysis, they compared temperature of the nose directly with that of the forehead. They reported a significant decrease in the nose temperature under the stress condition. On the other hand, forehead temperatures of participants were stable. Despite the use of contact-based sensor (not thermal imaging), the result showed greater potentials of the nasal thermal directional change as a stress indicator. This leads to the following studies using different experimental protocols and more importantly, using contact-free thermography. ROIs have also been extended to peripheral facial areas including peri-nasal regions, supraorbital and frontalis areas [30,87,90,99,114]

Using a high-resolution thermal imaging system, Or and Duffy [87] observed significant thermal changes of the nose in a negative direction during a driving simulation task (3min) and a real-field driving task (5min) in comparison with pre-driving sessions as the baseline. Both tasks were designed to induce a considerable amount of mental stressors. The authors confirmed there was no significant change in temperature of the forehead. More recently, the similar pattern was reported from studies using the Stroop color-word test [2]. During the Stroop test [111], participants are required to name the color of words while the meaning of each word was presented either incongruent or congruent. This test has been widely used as a means of producing cognitive load-induced mental stress in neuroscientific [112] and psychophysiological studies [15,99].

Engert *et al.* [30] have extended ROIs by observing temperature patterns from periorbital, perioral, corrugator and chin, as well as the nose and forehead regions. The authors conducted a study using the trier social stress test [67] that consists of a mock-up job interview and a difficult mental arithmetic task. The study reported a significant decrease in temperature not only on the nose tip, but also on chin and corrugator areas under the stressful condition. However, in a highly controlled experiment using a chin rest [114], such areas other than the nose tip did not respond to mental stressors. The study used a mental memory task and monitored the thermal directional change of 17 facial ROIs including the nose tip, chin and corrugator (outside an eye and around eyebrow). Such incongruent results may indicate limitations of the use of simple directionality-related metrics.
Given the limited capability of the single metric, a very recent study has proposed a richer set of metrics with the aim to capture multiple aspects of physiological variability in thermal changes [14]. Also, the study investigated this in two unconstrained experimental settings: mental arithmetic and physical assembly tasks in the real world. The study reported that the nose temperature decreases when participants carried on sedentary cognitive tasks, the thermal direction metric was not sufficiently sensitive to mental stress levels. Also, the authors [14] found the interaction effect of the different types of stress (mental overload and social pressure) on the nasal temperature change. Another study reported that cognitive load decreases the nose tip temperature in neutral valence whilst it does not in negative valence [88], confirming the interaction effects between different affective states.

### Fear and being startled

Promising findings from the thermal imaging studies on mental stress and workload have led to investigations on other affective states. Fear is the one that has been studied in association with the thermal directional response [21,68,70,85]. Kistler *et al.* [68] used a horror film to induce fear in participants. Compared with temperature measured prior to the task, participants showed temperature drops of up to 2°C on their fingertips. In another study [21], a sudden acoustic stimulus was used to elicit fear in people with post-traumatic stress disorder (PTSD). A temperature drop was generally observed over the whole face of participants. The nose tip showed the largest temperature decrease. Studies on facial temperatures of nonhuman primates have also supported the decrease of the nasal temperature as a measure of fear [70,85].
Facial temperature changes can also occur in startled states [89,109]. Shastri *et al.* [109] reported decreased temperatures of the upper lip and its surrounding regions of their participants during the presentations of unexpected startling sound effect (glass breaking sound). Similarly, the increase in temperature of the periorbital region of participants was observed after being exposed to a loud startling sound [89]. Another study using a similar protocol [35] reported no significant changes in temperature of the periorbital area, confirming the limitation of the simple direction metric.

### Sexual arousal, empathy, sympathy and love

It is clear that advantages of the non-contact measurement are attractive in its use to observe a person's affective state in social contexts [25,44,58,103]. Hahn *et al.* [44] studied sexual arousal during social interaction with interpersonal physical contact (i.e. physically touching one's body). The authors identified the closer relationship a participant has with his opposite gender partner, the higher increase of temperature the person showed on the nasal, periorbital areas and the lips. In another social context, it has been shown that compliment-induced blushes increase the cheek and forehead temperatures whilst they decrease temperatures on the periorbital area [58].
In terms of the socio-emotional development in children, the thermal directional changes were investigated in relation to empathy between mothers and their children in stressful situations [25]. The study reported a significant parallelism between their facial temperature changes (mother-child). Sympathy was also investigated. Ioannou et al. [59] used a sad film to induce sympathy in female participants. The authors found significant increase in temperature on the periorbital area, the forehead and the cheeks, the chin before crying and further increase in the temperatures during crying.
A recent study explored temperature changes associated with romantic love in couples [103]. Participants in the study were asked to take a look at pictures of their beloved partner for postulating love in memory and those of their friend during a baseline period. The study reported a general increment in participants' facial temperature when feeling love. In another work, pictures of infants and adult faces were presented to non-parent adults for exploring emotional valence [31]. The focus of the study was on investigating gender differences in the nose tip temperature changes. The study reported a much higher number of female participants showing increases in their nose temperature than male participants when infant faces were presented.

#### In the context of mockup crime and accident

Pavlidis *et al.* [91] conducted an experiment where participants were required to commit a mock crime and declare their innocence of the crime. When participants were interrogated, increases in temperature of their periorbital regions were observed. Although the finding was not statistically evaluated, the same pattern was observed in a study using a different crime-related protocol [98]. In another study where a mock unlucky accident was presented to children playing with a toy (pre-planned to be broken), guilty children showed a decrease in the nose tip temperature [57].

## 3.2 Automated affect recognition

As reviewed in the section above, studies have highlighted the capability of thermography in affective computing. However, the main focus has been on observing affect-related thermal directional changes of a certain ROI through manual annotation, not using such signatures to recognize a person's affective states automatically. In this section, we review studies on automatic affect recognition using thermal imaging.

Nhan and Chau [86] proposed an affect recognition system with a heavyweight, high-resolution system. They focused on the inference of arousal (unexcited-excited) and valance (unpleasant-pleasant) states (i.e. high or low). The authors set five ROIs: left, right supraorbital, left, right periorbital, and the nose. Participants were given a set of pictures from the international affective picture system (IAPS) [71]. With thermography, the authors used additional contact-based physiological sensors to measure heart rate and respiration rate. To avoid the complexity of automatically tracking each ROI, a dot stick was attached to the top of their participants' forehead. 78 values from each ROI (using basic statistical and customized functions), 10 correlation coefficients between each ROI pair (i.e. ${}_5C_2$ – *combination*) and average heart and respiratory rates (from the contact-based sensors) were used as machine learning features. They employed the linear discriminant analysis (LDA) as a binary classifier. In k-fold cross validation (k=10), the system achieved an accuracy of approximately 80% in discriminating high arousal state from a baseline state, and high valence state from a baseline state, respectively. They reported lower-chance level accuracies for the case of classifying high arousal versus low arousal, and high valence versus low valence.

This approach [86] has been extended by a couple of studies. A support vector machine (SVM)-based system with features from the supraorbital, periorbital and the forehead, the nose and mouth achieved the accuracy of 63.5% in the binary detection of sadness and happiness (with k-fold cross validation) [72]. A very recent study proposed a similar approach using such features with the LDA to recognize five emotions: disgust, fear, happiness, sadness and surprise [41]. In k-fold cross validation (k=3), the approach achieved the lowest accuracy of 74.7% for sadness and the highest of 89.88% for disgust. Interestingly, Khan *et al.* [65] proposed a segmentation of a person's facial thermal image into a number of local areas (e.g. 75 local squared areas) to extract an extensive number of thermal features. In recognizing happiness and sadness on the data collected from the IAPS task, the LDA-based system achieved accuracies of 73.7% and 68.4% for each case (the type of cross validation was not reported). Similar techniques were applied to the detection of stress [55] and thermal discomfort [9] with both reporting the accuracy of circa 90%. However, the stress detection study [55] did not cross validate properly and the discomfort study used a non-standard cross-validation method: 42 fold leave-one-out cross validation with 14 participants.

Automatic feature learning is a power centralised in most successfully and widely used deep learning approaches in computer vision and pattern recognition [73]. This helps automatically find good features during the machine learning process. As even carefully hand engineered-feature extractors could fail to generalise to unseen data sets [73], this could be a potential solution to improve our understandings of physiological and behavioural patterns in relation to a person's affective states. Given this, very recent works have pushed the boundaries by proposing end-to-end learning systems with new representations of physiological thermal signatures [12,15].

Cho et al. [12] proposed an end-to-end stress detection system with a representation of respiratory thermal signatures (respiration variability spectrogram) that condense complex stress-related respiration variability with two-dimensional data as shown in Figure 8. This helps automatic feature learning methods (e.g. deep convolutional neural network, CNN) search for informative features. The built respiration-based stress recognition system has achieved state-of-the-art performance despite the use of only the breathing signature. The reported accuracy was 84.59% in discriminating two stress levels. The accuracy was achieved from a k-fold leave-one-subject-out (LOSO) cross validation. As illustrated in Figure 9, at every fold, all data sequences from the participants except for one were used to train the networks, and the data from the left-out participant was used for testing. This is known to be difficult to achieve a high accuracy given its nature to test the ability to generalize to unseen participants' physiological data [12,15,29,53]. Using the released dataset and system (http://youngjuncho.com/datasets), we tested the approach using the k-fold cross validation (k=10) and the accuracy was 99.7%.

Likewise, another study proposed an end-to-end learning framework for the instant stress detection that combines two different types of cardiovascular signatures: vasoconstriction-induced nasal temperature and blood volume pulse [15]. By focusing on capturing physiological variability, the system achieved good performance for inferring mental stress levels from a LOSO cross validation (78.33% accuracy).

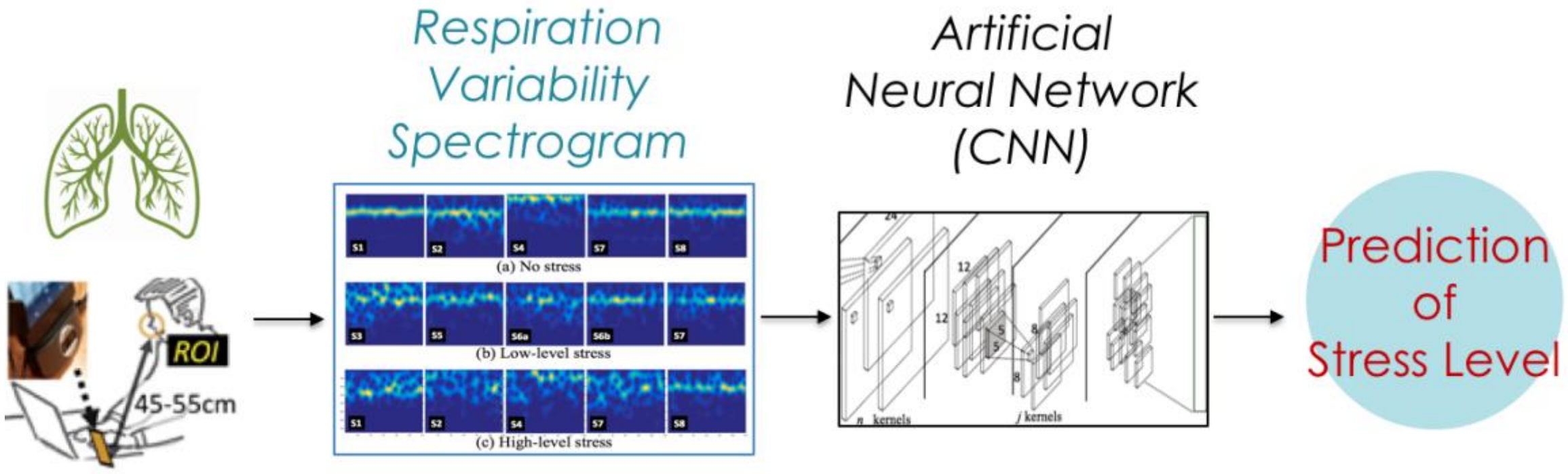

**Figure 8.** An end-to-end learning framework for automatic stress detection proposed in [12] consisting of respiration variability spectrogram and artificial neural networks (adapted from [12]).

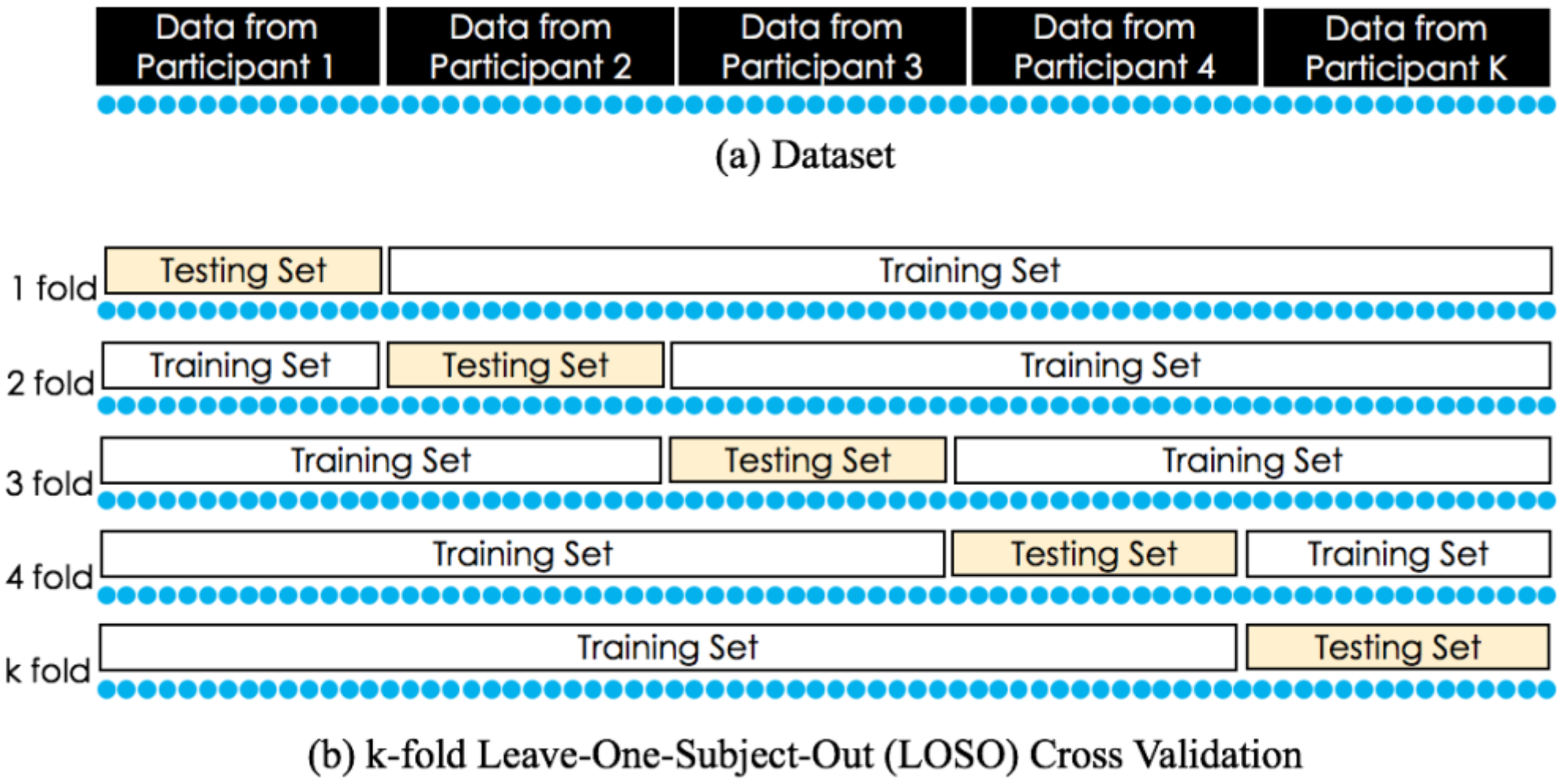

**Figure 9.** k-fold leave-one-subject-out (LOSO) cross validation method: (a) dataset collected from k participants, (b) splitting the dataset into training and testing sets for each fold.

## 3.3 A summary of experimental protocol and data analysis

Table 3 summarizes reviewed approaches for thermography-based affective computing. We elaborate the key information of study protocols used in the literature (e.g. task types and baseline) in Table 3, providing an insight for building experimental protocols. As discussed, the directional change has been used as a metric in general. The averaging has been dominantly used to represent temperatures of a ROI on thermal images. Furthermore, the majority of existing studies (reviewed in both Section 2 and 3) were conducted in systematic constrained settings that require stable environmental (ambient) temperature and low-level motion artefacts whilst a couple of recent works have explored how to bring thermal imaging to unconstrained real world, ubiquitous settings [12–16]. Although a few studies have explored automatic ROI tracking (also called motion tracking) methods, the tracking performance in varying thermal range scenes has still remained less explored [12,16].

**Table 3.** A summary of thermography-based affective computing studies: experimental protocols, required constraints, explored ROIs and metrics.

| **Affective States** | **Author** | **Experimental protocols** | | | | **Analysis** | |
|---|---|---|---|---|---|---|---|
| | | Participants | Task | Baseline | Requirements | ROIs | Metrics or Features[2] |
| **Mental Stress** | Genno et al. (1997) | 117 healthy adults | Self-made tracking task (11mins) | Rest (5 mins) | Stable room temperature | Nasal and forehead | Difference between mean temperature (from two points) & Temperature difference between data from two ROIs |
| | Veltman and Vos (2005) | 8 adults | Continuous Memory Task (3mins) | Rest (3mins) | Chin rest used | Nasal, cheek, chin, lips, eyes, forehead | Difference between mean temperature (from two points) |
| | Or and Duffy (2007) | 33 healthy licensed drivers | Car-driving simulator tests (3mins) | Rest (3mins) | Stable room temperature | Nasal and forehead | Difference between mean temperature (from two points) |
| | Pavlidis et al. (2012) | 17 healthy adults | Laparoscopic drill training (4mins) | Natural landscape | Not reported (but, see Figure 4) | Perinasal and Finger | Difference between mean temperature at two temporal points (on sweat glands) |
| | Engert et al. (2014) | 15 male adults | 1) Cold Pressor Test, 2) Trier Social Stress Test | Rest (5mins) | Stable room temperature & Visual inspection | Chin, Corrugator, Forehead, Finger, Nose, Periorbital, Perioral | Mean slope |
| | Cho et al. (2017) | 8 healthy adults | 1) Stroop test, 2) Stressful arithmetic task | Rest (5mins) | Unconstrained – No room temperature control & No mobility control | Nostril | Respiration rate & respiration variability spectrogram |
| | Cho et al. (2019) | 17 healthy adults | | None | | Nose tip | Thermal variability sequence & blood volume pulse sequence |

[2] Although metrics and features are used as synonym in some literature, in this article, metric is used to describe a measure to quantify physiological symptoms and feature is used to describe specific feature inputs to, or analyzed in, machine learning models.

| | | | | | | | |
|---|---|---|---|---|---|---|---|
| | | | | | | | from finger photoplethysmography |
| | Cho et al. (2019) | 15 healthy adults | 1) Stressful arithmetic task<br>2) real world furniture assembly task | Rest (5mins) | | Nose tip | Difference between mean temperature,<br>Slope of thermal variable signal,<br>Standard deviation of successive differences of thermal variable signal,<br>Standard deviation of thermal variable signal |
| **Cognitive Load** | Pinti et al. (2015) | 9 healthy participants (for the full experiment) | Block-based cognitive load task | Rest (15sec) | Stable room temperature and no ventilation | Nose tip | Wavelet for the very low frequency (VLF) range (0.008-0.08Hz) |
| | Abdelrahman et al. (2017) | 24 Egyptian and Canadian participants | Stroop Test | Rest (10mins) | Stable room temperature & Keeping participants still and facing a thermal camera | Nose tip & forehead | Difference between mean temperature (from two points) |
| **Fear** | Kistler et al. (1998) | 20 healthy Caucasian adults | Watching horror movie scene (3mins 45s) | Rest (2mins) | Stable room temperature | Finger tip | Difference between mean temperature (from two points) |
| | Di Giacinto et al. (2014) | 10 post traumatic stress disorder patient, 10 healthy adults | Image presentation with 80-db white noise burst (6mins) | Image presentation with no sound (6mins) | Stable room temperature &without ventilation<br>& a dot stick was used | Nasal and whole face | Difference between mean temperature (from two points) |
| **Deception** | Pollina et al. (2006) | 30 healthy adults | Committing a mock crime | not specified | Stable room temperature & Visual inspection | Left and right hemiface for each eye area | Difference between mean temperature (maximum and minimum temperature image) |
| **Startle reaction** | Pavlidis et al. (2001) | 6 healthy participants | A startle stimulus using a sudden loud noise (60dB) | Sitting in the dark room (10 mins) | Stable room temperature | Periorbital, Cheeks, Neck, Nasal, chin region | Average pixel values (relative thermal change) |
| | Shastri et al. (2012) | 18 healthy participants | Presentation of unexpected startle sounds (e.g. glass breaking sound). | not specified | Not reported (but, a similar environment to the one in Figure 2.1) | Upper lip and surrounds | Difference between mean temperature (from two points) & **wavelet**-based metrics |
| | Gane et al. (2011) | 11 adults | 102 dB Auditory startle stimulus (instantaneous (30s) x 100 repetitions) | Rest (1min) | Manually adjusting thermal imaging parameters to room temperature (an ambient | Periorbital areas | Range, mean, variance, skewness, kurtosis, entropy, normality, stationarity<br>of<br>2 second mean temperature timeseries |
| | | | | | temperature sensor was used) | | |
| **Sexual arousal** | Hahn et al., (2012) | 23 healthy adults | Facial contact with an opposite-sex experimenter (15mins) | Prior to facial contact | Keeping participants from moving | Nasal and periorbital areas and lips | Difference between mean temperature (from two points) |
| **Maternal empathy** | Ebisch et al. (2012) | 12 mothers and their children | Eliciting stressful situations based on the "mishap paradigm" | Prior to task (10-20mins) | Stable room temperature | Nose tip, maxillary regions | Difference between mean temperature (from two points) |
| **Emotional valence** | Esposito et al. (2015) | 38 non-parent adults (19 Females) | Pictures of infant and female faces given | Presentation of a neutral screen (10sec) | Climate chamber with a stable room temperature | Nose tip | Difference between mean temperature (from two points) |
| **Love** | Salazar-López et al. (2015) | 12 couples | Presentation of pictures of the beloved partner (5mins) | Presentation of pictures of friends (5mins) | Keeping participants from moving | Whole face, hand | Difference between mean temperature (from two points) |
| **Positive-negative emotions** | Nhan & Chau (2010) | 12 healthy adults | Presentation of pictures selected from the international affective picture system (IAPS) | Rest | Stable room temperature & a dot tracker used | left, right supra-orbital, periorbital, nasal regions | 390 features (5ROIs x 78) and 10 features (correlation coefficients from pairs of ROIs) (*for classification) |
| | Salazar-López et al. (2015) | 120 university students | Presentation of pictures selected from the international affective picture system (IAPS) (several seconds) | Rest (10mins) | Keeping participants from moving | Nose tip, forehead, orofacial area and cheeks and the whole face region | Difference between mean temperature (from two points) |
| **Facial expressions and Facial Action Coding Systems (FACS)** | Khan et al. (2004), (2016) | 16 undergraduates (2004), 19 participants (2016) | Presentation of pictures selected from the international affective picture system (IAPS) (several seconds) | Rest (20mins) | Stable room temperature | Self-defined facial thermal feature points | Covariance metrics and maximum temperature timeseries from 75 self-selected ROIs on a face (*for classification) |
| | Wang et al. (2010) | 215 healthy students | Presentation of film clips | Not reported | Unconstrained room environments | Forehead, nose, lips, cheek | Feature selection using PCA, PCA+LDA, AAM, AAM+LDA, etc. (*for classification) |
| | Jarlier et al. (2011) | 4 trained and certified FACS coders | Activating different facial action units | Rest (15mins) | Keeping participants' heads immobilised with a head fixation system | Facial action units | Topography of thermal changes based on speeds and intensities of muscle contraction (*for classification) |

## 3.4 A summary of specifications of thermal imaging systems used in the literature

Table 4 summarizes the specifications (resolutions, sensitivity) of thermographic systems used in the literature discussed in Section 2 and Section 3. Also, it describes actual specifications required for the thermal analysis. The majority of studies have employed very heavy and expensive thermographic imaging systems (e.g. Figure 1b, Figure 4) which are inaccessible to the general public. In addition, installation difficulty has confined ROIs mostly to the facial regions, possibly interfering with one's vision, causing distraction, and more importantly, prohibiting this from being carried out in unconstrained, mobile settings.

Interestingly, many works have required lower spatial resolutions and sampling rates than these of their systems (highlighted in bold with brackets, Table 4). Very low sampling rates or static data have been required for capturing thermal directional changes; for example, the sampling rate of 340Hz was down-sampled to 0.2Hz in [114]. Respiratory rate can also be extracted at a lower sampling rate of 5Hz [75]. The spatial resolution of 640x512 was also down-sampled to 100x100 in [109].

**Table 4.** Specification requirements for thermography-based physiological measurements and affective computing.

| **Author** | **Metrics** | **Affective States** | **Specification of Thermal Camera** | | |
|---|---|---|---|---|---|
| | | | Spatial Resolution | Sampling Rate | Thermal Sensitivity |
| **Fei & Pavlidis, (2010)** | *respiratory rate* | Not explored | 640x512 | 55Hz **(10Hz)** | 0.025℃ |
| **Murthy *et al.* (2004)** | | | 640x512 | 126Hz **(31Hz)** | 0.025℃ |
| **Pereira *et al.* (2015)** | | | 1024x768 | 30Hz | 0.05℃ |
| **Cho *et al.* (2017a)*** | *respiratory rate & variability* | | 160x120 | <9Hz | 0.07℃ |
| **Lewis *et al.* (2011)** | *respiratory rate & relative tidal volume* | | 320x240 | <29Hz **(5Hz)** | 0.08℃ |
| | | | 640x510 | 126Hz **(30Hz)** | 0.02℃ |
| **Shastri *et al.* (2009)** | *skin-conductance-related raw signal* | | 640x512 | 126 Hz | 0.025℃ |
| **Abdelrahman *et al.* (2017)** | *thermal directional change* | cognitive load | 160x120 | 120Hz | 0.08℃ |
| **Di Giacinto *et al.* (2014)** | | fear | 320x240 | 60Hz **(10Hz)** | 0.02℃ |
| **Ebisch *et al.* (2012)** | | empathy | 320x240 | 50Hz **(1Hz)** | 0.02℃ |
| **Engert *et al.* (2014)** | | mental stress | 320x240 | 50Hz **(5Hz)** | 0.02 ℃ |
| **Gane *et al.* (2011)** | | startled state | 640x480 | 120 Hz **(15Hz)** | 0.01℃ |
| **Hahn *et al.* (2012)** | | sexual arousal | 160x120 | 1/75 Hz | 0.08 ℃ |
| **Kistler *et al.* (1998)** | | fear | 280x90 | 6.25Hz **(0.1Hz)** | Not reported |
| **Nakayama *et al.* (2005), Kuraoka *et al.* (2011)** | | fear | 254x238 | 0.1Hz | 0.1℃ |
| **Or and Duffy (2007)** | | mental stress | 320x240 | 60Hz **(Static image)** | 0.08℃ |
| **Pollina *et al.* (2006)** | | deception-related anxiety | 256x256 **(256x150)** | 30Hz | 0.1 ℃ |
| **Salazar-López *et al.* (2015)** | | romantic love | 320x240 | 60Hz **(Static image)** | 0.07 ℃ (**0.1 °C – 0.5 °C)** |
| **Shastri *et al.* (2012)** | | startled state | 640x512 (**100x100)** | 126 Hz **(25Hz)** | 0.025℃ |
| **Veltman and Vos (2005)** | | mental stress | 320x256 | 340Hz (**0.2Hz)** | 0.07℃ |
| **Pinti et al. (2015)** | Wavelet phase coherence (frequency domain analysis) | cognitive load | 640x480 | 30Hz **(2Hz)** | 0.03℃ |
| **Pavlidis *et al.* (2012)** | *thermal directional change* & *skin-conductance-related raw signal* | mental stress | 640x512 | 126Hz **(25Hz)** | 0.025℃ |
| **Cho *et al.* (2017b)*** | *respiratory rate & variability* | | 160x120 | <9Hz | 0.07℃ |
| **Cho *et al.* (2019a,b)*** | *Thermal directional change & Thermal variability* | | 160x120 | <9Hz | 0.07℃ |

**() Bold within brackets:** Studies requiring lower spatial resolutions and sampling rates than specifications of the systems
* shaded row: Mobile thermal imaging

# 4 Challenges and opportunities

The ability to monitor physiological vital signs and affective states is becoming more important in human computer interaction [4,56,96]. Such functions are being integrated into systems aimed to address real world problems (e.g. health monitoring, educational support, entertainment) in both research and commercial products. In many of these situations, mobile approaches are required to enable the users to undertake their activities. Mobile thermal imaging has greater potentials in its use as a multimodal physiological and affect sensor for ubiquitous real-world settings. In this section, we discuss existing challenges and research opportunities emerging from the literature.

## 4.1 Beyond thermal directional changes: Diversifying metrics for real world applications

As reviewed, most of existing studies have focused on the simple metric designed to observe a binary direction of thermal changes (i.e. temperature decrease, increase) in association with a person's affective states. Despite findings opening up exciting prospects for affective computing, they have often shown incongruent results (e.g. chin temperature change in response to stressors: decrease in [30] whilst no change in [114]). This indicates that a single discrete metric is susceptible to diverse factors due to the complex physiological mechanisms (e.g. a voluntary facial muscular action can change the temperature [102]). Also, there must be interaction effects between affective states [14,88]. All of these have the potential to induce temperature variability rather than just a consistent drop or increase in temperature. For example, continuous variations in the nose tip temperature were observed during 2hours simulated driving tasks [22]. The simple metric could possibly lose important information of psychophysiological reactions. Hence, it is required to diversify metrics.

To bring variety to thermal metrics, extreme robustness of physiological measurements is required. State-of-the-art methods for extracting physiological thermal signatures can produce rich, less noisy physiological signals [14,16]. By richness, we are referring to the complexity of physiological signatures. Capturing complex variations in physiological patterns can help diversify metrics for variability beyond single metric values. Although a few of recent studies have proposed new metrics to capture successive differences of thermal variable signals [15] and very low frequency components using wavelet phase coherence [10], further efforts are still needed to avoid collapsing the complex physiological phenomena into a single metric.

Indeed, the ability to capture physiological variability has been extremely important in psychophysiology beyond thermal imaging studies. For example, given the fight-or-flight responses to stressors [32], stressful situations induce irregular breathing patterns, not just increasing the breathing rate [12,43]. With a representation of physiological variability, automatic feature learning techniques (also called representation learning [73]) can also help capture the variability. As discussed, artificial neural networks can be of use to extract the variability-related information from respiration variability spectrograms that represent breathing signals [12]. It is expected that it is possible to capture informative variability from other types of physiological thermal signatures.

## 4.2 ROI tracking and Ambient temperature effect

Recent advances in thermographic systems have shrunk their size, weight and cost, to the point where it is possible to support mobile thermal imaging. This is not affected by lighting conditions (e.g. night illumination incapacitating remote PPG [115]). It enables to measure temperature anywhere in mobile, temperature varying situations.

In capacitating mobile thermal imaging to function in real-world settings, however, we must ensure to track a ROI on thermal images extremely reliably. Even in controlled laboratory settings, a person’s movement makes physiological thermal signatures noisy (e.g. a sudden change in a person’s head direction produces noise [16]) and a ROI tracking error could deteriorate their quality [109,113]. The situation gets even worse if we bring thermal imaging to in-the-wild situations beyond stable temperature environments. This is because of dynamic changes in ambient temperature impacting the skin temperature and resulting in inconsistent thermal shapes. This makes it much harder to track ROIs automatically.

Despite recent efforts to handle ambient temperature effects [16], the conducted experiments (e.g. outdoor breathing exercise, physical activities) cannot cover all possible scenarios such as a swimming pool, the seaside where humidity affects temperature measurements and different geographical zones and extremely hot days and different climate types. A simple solution to improve ROI trackers could be to use RGB camera or other remote tracking systems together with a thermal camera as in facial expression studies [6,51]; however, it could restrain its portability, flexibility and increase computational cost (this further limits real time tracking performance). Hence, more attention needs to be paid to developing computational methods to support automatic ROI tracking on thermal images in entirely mobile, ubiquitous situations.

## 4.3 Evaluation, Datasets and Toolkits

The evaluation of thermal imaging-based affective computing systems is complex given its nature with different focuses and a broad range of contexts and experimental protocols. On the other hand, the systematic evaluation of certain physiological thermal signatures, such as respiration, perspiration and cardiac pulse, can be relatively straightforward given standard evaluation tools and methods in computational physiology. As for the vasoconstriction-related thermal signature, the quality of physiological signals was not systematically evaluated given no standard reference measurement.

Some authors, however, adopted nonstandard evaluation metrics (i.e. not used in computational physiology) for cardiac pulse and breathing rates, potentially leading to misconceptions. An example is the CAND [34,36,48] that tends to produce extremely higher values (e.g. "CAND accuracy=92.46%" in [48] turns out to be a weak correlation of "r=0.58" as pointed out in Section 2.2). Even a highly cited work [36] used this metric. This has led to the emergence of poor applications which implement these systems without a careful review, such as an assistive robot with thermal imaging-based heart rate monitoring [19] reporting poor results. Likewise, some works on automatic affect recognition reported extremely strong performance with the use of nonstandard cross-validation methods. In the classification on human physiological data, LOSO cross validation has been strongly recommended to avoid artificially high results due to training and testing a machine learning model on temporally adjacent samples [15,29,53].

Furthermore, the evaluation of the systems was done mostly on healthy participants. Given a few of studies reporting differences across different groups (gender [31] and people with mental impairment [95]), evaluation needs to be taken on a broad range of participants groups. With the use of rigorous evaluation methods, this could help formulate more standard and powerful frameworks that can be used in the real-world environments.

Lastly, datasets and toolkits in this field of science have been limited. Only a couple of thermal imaging datasets ([12,16]) are publicly available. Free software for the thermal analysis (e.g. FLIR Tools) provides a limited functionality. To foster this promising research across communities, there is a need to make them available to the public. Following this review, we release an open-source toolkit for Thermal Imaging-based Physiological and Affective computing (TIPA)[3].

# 5 Summary

In this article, we discussed studies and methods for physiological and affective computing through thermal imaging. Thermal imaging of the human skin can be of use in monitoring a person's physiological responses: cardiovascular, respiratory and perspiratory, muscular responses. In particular, vasoconstriction, vasodilation and sweat gland activation can lead to a directional change in temperature (rise or drop). Given the simplicity, the majority of the studies have focused on finding the relationship between the directional change and one's affective states. Affective states explored in the body of work span from mental stress, fear, startle, to love and maternal empathy. Overall, this article has provided a current pipeline of this research area with four components: i) ROI (where to monitor), ii) physiological thermal signature (what to observe and how to track), iii) metrics (how to quantify), and iv) affective states. In doing so, we have paid particular attention to computational methods.

In this survey, we have highlighted promising beneficiaries of thermal imaging-based approaches. Thermography itself does not necessarily require the sensors to be worn directly on the skin. It raises less privacy concerns and is less sensitive to ambient lighting conditions. It can benefit general healthcare and fitness sectors by providing contactless, simultaneous measurements of multiple vital signs to specific user groups. For example, people with chronic pain who tend to reduce the number of clothes and/or devices that touch their body.

This article has also paid attention to mobile thermal imaging as a possible solution for real-world applications. The use of small, light-weight, and more importantly, low-cost sensing devices makes itself more feasible and practical. This provides a new pathway, addressing limitations of static thermal imaging in highly constrained settings - which has been dominantly used in this area - and in turn providing practical solutions that work in HCI in-situ and in mobile contexts. All in all, we believe this article can act as a guideline to, and foster, the emerging physiological and affective computing methods in the research community.

[3] TIPA opensource toolkit: http://youngjun.cho/TIPA